\documentclass[12pt]{article}

\usepackage{amsmath,amssymb}
\usepackage{graphicx}

\newcommand{\reseteqnum}{\setcounter{equation}{0}}

\begin{document}

\title{
\hfill
\parbox{4cm}{\normalsize
hep-th/0508248}\\
\vspace{2cm}
A relation between moduli space of D-branes on orbifolds and Ising model
\vspace{2cm}}
\author{Tomomi Muto
\vspace{1.0cm}\\
{\normalsize Tokyo Metropolitan Institute of Technology}\\
{\normalsize Asahigaoka 6-6, Hino-shi, Tokyo 191-0065, Japan}\\
{\normalsize tmutoh@cc.tmit.ac.jp}
}
\date{\normalsize}
\maketitle

\vspace{1.0cm}

\begin{abstract}
We study D-branes transverse to an abelian orbifold ${\bf C}^3/{\mbox{\boldmath $Z$}}_n \times {\mbox{\boldmath $Z$}}_n$.
The moduli space of the gauge theory on the D-branes is analyzed by combinatorial calculation based on toric geometry.
It is shown that the calculation
is related to a problem to count the number of ground states of an antiferromagnetic Ising model.
The lattice on which the Ising model is defined is a triangular one defined on the McKay quiver of the orbifold.
\end{abstract}

\newpage
\section{Introduction}

The moduli space of the gauge theory on D-branes transverse to an orbifold ${\bf C}^3/\Gamma$
has been investigated based on the procedure proposed by Douglas, Greene and Morrison \cite{DGM, Muto, BGLP, FHH, Sarkar}.
In this approach the geometry of the moduli space,
which is defined by F-flatness and D-flatness conditions,
is analyzed in the language of toric variety
and the result is described by a cone in a integer lattice ${\mbox{\boldmath $Z$}}^c$.
The dimension $c$ is determined by a calculation of a dual cone
which is necessary to make the F-flatness and D-flatness conditions on the same footing.
To find the dual cone we must carry out a combinatorial algorithm based on the linear programming,
but it is difficult to obtain an explicit result in general.
By inspecting closely the calculation of the dual cone, however,
one can see that it does not fully use the structure of the gauge theory.

One of the notable feature of the gauge theory is that it is a quiver gauge theory \cite{DM, JM}.
Information on the gauge theory such as F-flatness and D-flatness conditions are encoded in
the McKay quiver of the group $\Gamma$ \cite{McKay, Reid}.
So if we fully make use of the information as a quiver gauge theory,
the calculation of the dual cone will become easier.
Based on this observation
we show that the combinatorial calculation of the dual cone is equivalent to a problem
of counting the number of ground states of a certain Ising model on a lattice defined on the McKay quiver of the orbifold.
In this paper we consider the orbifold ${\bf C}^3/{\mbox{\boldmath $Z$}}_n \times {\mbox{\boldmath $Z$}}_n$,
in which case the Ising model is an antiferromagnetic one on a triangular lattice with certain boundary conditions.

In section 2 we review the gauge theory on
one D3-brane transverse to a quotient singularity ${\bf C}^3/\Gamma$
with $\Gamma$ an abelian subgroup of $SU(3)$.
We also recapitulate the procedure \cite{DGM} for the analysis of the moduli space $\cal M$ of the gauge theory.
In section 3, we investigate the orbifold ${\bf C}^3/{\mbox{\boldmath $Z$}}_n \times {\mbox{\boldmath $Z$}}_n$.
We show that the calculation of the dual cone can be replaced by a calculation of ground states of
an antiferromagnetic Ising model on a triangular lattice constructed from the McKay quiver.
In section 4, we present a systematic way to count the number of the ground states
by using transfer matrices of the Ising model.
We also generalize the analysis to the orbifold ${\bf C}^3/{\mbox{\boldmath $Z$}}_n \times {\mbox{\boldmath $Z$}}_m$.

This paper is based on a talk "A relation between quiver variety and Ising model"
at the autumn meeting of the Physical Society of Japan, September 2003.

\section{The moduli space for one D-brane transverse to an abelian quotient singularity ${\bf C}^3/\Gamma$}
\reseteqnum

In this section, we briefly review the gauge theory on
one D3-brane transverse to a quotient singularity ${\bf C}^3/\Gamma$
with $\Gamma$ an abelian subgroup of $SU(3)$.

\subsection{The gauge theory on the D-brane and the McKay quiver}

The gauge theory on one D3-brane transverse to a quotient singularity ${\bf C}^3/\Gamma$
is obtained by considering $|\Gamma|$ D3-branes on ${\bf C}^3$, and then projecting to ${\bf C}^3/\Gamma$.
The worldvolume theory on $|\Gamma|$ D3-branes on ${\bf C}^3$ is a theory with $U(|\Gamma|)$ gauge symmetry,
containing a vector multiplet $A_a$ $(a=0, \cdots, 3)$
and three chiral multiplets $X^\mu$ ($\mu =1, 2, 3$) in the adjoint representation of the gauge group.
The components of the gauge field $A_a$ that survive the projection are those which satisfy
\begin{equation}
\gamma (g) A_a \gamma (g)^{-1} = A_a
\label{eq:projA}
\end{equation}
where $g \in \Gamma$, and $\gamma(g)$ is the regular representation of $\Gamma$.
For an abelian group $\Gamma$, the regular representations takes the form
\begin{equation}
\gamma (g)=\oplus_{v=0}^{|\Gamma|-1} R_v (g)
\end{equation}
where $R_v$ are one-dimensional irreducible representations of $\Gamma$.
The condition (\ref{eq:projA}) then implies that all elements of $A_\alpha$ must be zero except for diagonal components
and hence the gauge group is isomorphic with $U(1)^{|\Gamma|}$.
Note that the overall $U(1)$ acts trivially on all fields,
so the effective gauge symmetry is $U(1)^{|\Gamma|-1}$.
The surviving components of $X^\mu$ are those which satisfy
\begin{equation}
\gamma (g) X^\mu \gamma (g)^{-1} = R(g)^\mu_\nu X^\nu
\end{equation}
where $R(g)$ is the three-dimensional representation
defining the action of $\Gamma$ on ${\bf C}^3$.
We choose the three-dimensional representation of the form
\begin{equation}
R(g)=R_{v_1} (g) \oplus R_{v_2} (g) \oplus R_{v_3} (g)
\label{eq:repR}
\end{equation}
with $R_{v_1} \otimes R_{v_2} \otimes R_{v_3}$ to be the trivial representation $R_0$.
All elements of $X^\mu$ must be zero except for
the $3|\Gamma|$ components $X^1_{v \; v+v_1}, X^2_{v \; v+v_2}, X^3_{v \; v+v_3}$,
which we denote as
\begin{equation}
x_v = X^1_{v \; v+v_1},\quad
y_v = X^2_{v \; v+v_2},\quad
z_v = X^3_{v \; v+v_3}.
\end{equation}
Charges of the $v$-th $U(1)$ gauge symmetry of $x_w, y_w, z_w$ are given by
\begin{eqnarray}
q_v (x_w) &=& \delta_{w,v+v_1}-\delta_{w, v}, \nonumber\\
q_v (y_w) &=& \delta_{w,v+v_2}-\delta_{w, v}, \label{eq:charge}\\
q_v (z_w) &=& \delta_{w,v+v_3}-\delta_{w, v} \nonumber
\end{eqnarray}
respectively.

Gauge group and matter content of the world volume theory on the D-brane are specified by a McKay quiver.
The McKay quiver
is a graph with a set of nodes $V$
and a set of arrows $A$ connecting nodes.
Each node, which has one to one correspondence with the one-dimensional irreducible representation $R_v$
of $\Gamma$,
gives rise to a factor $U(1)$ of the gauge group.
The set of arrows of the quiver
is determined by the following irreducible decomposition
\begin{equation}
R \otimes R_v =\oplus n_{vw} R_w.
\end{equation}
Non-negative integer $n_{vw}$ is the number of arrows from node $v$ to node $w$.
For the representation $R(g)$ in (\ref{eq:repR}),
the decomposition takes the form
\begin{equation}
R \otimes R_v = R_{v+v_1} (g) \oplus R_{v+v_2} (g) \oplus R_{v+v_3} (g),
\end{equation}
thus we have three arrows ($a_v, b_v, c_v$) leaving the node $R_v$:
\begin{eqnarray}
a_v &:& R_v \rightarrow R_{v+v_1}, \nonumber \\
b_v &:& R_v \rightarrow R_{v+v_2}, \\
c_v &:& R_v \rightarrow R_{v+v_3}. \nonumber
\end{eqnarray}
Note that arrows $a_v, b_v, c_v$ correspond to the fields $x_v, y_v, z_v$, respectively.

\subsection{The moduli space of the gauge theory}

The classical moduli space $\cal M$ of the gauge theory is obtained
by imposing the F-flatness and D-flatness conditions of the theory
and further dividing by the gauge group $U(1)^{|\Gamma|-1}$.
The F-flatness conditions come from the superpotential
\begin{equation}
W={\rm Tr} \, [X^1, X^2] X^3
\end{equation}
whose minimization yields the matrix form of the F-term equations
\begin{equation}
[X^\mu, X^\nu]=0.
\end{equation}
In components, the equation takes the form,
\begin{equation}
x_v y_{v+v_1} = y_v x_{v+v_2}, \quad
y_v z_{v+v_2} = z_v y_{v+v_3}, \quad
z_v x_{v+v_3} = x_v  z_{v+v_1}.
\label{eq:F}
\end{equation}
Note that the number of independent F-term equations is $2|\Gamma|-2$,
and hence these equations define a $|\Gamma|+2$ dimensional complex subvariety $\cal Z$.
Since $\cal Z$ has the structure of an algebraic variety defined by a collection of monomial relations,
it is an affine toric variety and admits an alternative presentation as a holomorphic quotient.
If we change notation temporarily as
\begin{equation}
(u_1, \cdots, u_{3|\Gamma|})=(x_0, \cdots, x_{|\Gamma|-1}, y_0, \cdots, y_{|\Gamma|-1}, z_0, \cdots, z_{|\Gamma|-1}),
\end{equation}
we can solve the equations and express $u_l$
in terms of $|\Gamma|+2$ parameters $v_\alpha$ ($\quad \alpha=1, \cdots, |\Gamma|+2$);
\begin{equation}
u_l = \prod_\alpha v_\alpha^{K_{l \alpha}}. 
\label{eq:uv}
\end{equation}
The rows of the $3|\Gamma| \times (|\Gamma|+2)$ matrix $K = (K_{l \alpha})$ are vectors in the lattice $M={\bf Z}^{|\Gamma|+2}$,
and define the edges of a cone $\sigma$.
To describe $\cal Z$ as a holomorphic quotient,
we define a dual cone $\hat \sigma$ as follows;
\begin{equation}
\hat \sigma =
\{ \mbox{\boldmath $n$} \in {\bf R}^{|\Gamma|+2} \, | \, \mbox{\boldmath $m$} \cdot \mbox{\boldmath $n$} \geq 0 \quad
{\rm for} \quad \forall \mbox{\boldmath $m$} \in \sigma \}.
\label{eq:dual}
\end{equation}
We introduce a matrix $T$ such that its columns span the dual cone $\hat \sigma$.
The dimension of the matrix $T$ is $(|\Gamma|+2) \times c$ where $c$ is the number of edges of $\hat \sigma$.
Each column of $T$ corresponds to a homogeneous coordinate
$p_\lambda$ ($\lambda=1, \cdots, c$) of the holomorphic quotient description of $\cal Z$,
\begin{equation}
{\cal Z} = {\bf C}^c/({\bf C}^*)^{c-|\Gamma|-2}.
\label{eq:holomorphic}
\end{equation}
The action of $({\bf C}^*)^{c-|\Gamma|-2}$ on $p_\alpha$ is determined by a $(c-|\Gamma|-2) \times c$ matrix $Q_F$
defined by $Q_F=({\rm Ker} T)^t$.

Note that
the variables $v_\alpha$ are solved in terms of $p_\lambda$;
\begin{equation}
v_\alpha = \prod_\lambda \, p_\lambda^{T_{\alpha \lambda}}.
\end{equation}
Combining these equations with (\ref{eq:uv}), we obtain
the relationship between $u_l$ and $p_\lambda$
\begin{equation}
u_l = \prod_\lambda \, p_\lambda^{K_{l \alpha} T_{\alpha \lambda}}.
\label{eq:up}
\end{equation}
Due to the definition of the dual cone,
the exponent $K_{l \alpha} T_{\alpha \lambda} = (KT)_{l \lambda}$
takes value in ${\mbox{\boldmath $Z$}}_{\ge 0}$.

To obtain the vacuum moduli space ${\cal M}$ of the quiver 
gauge theory, one must further impose $|\Gamma|-1$ D-flatness conditions
that exist in the quiver gauge theory from the beginning.
The D-term equation for the $v$-th $U(1)$ group is given by
\begin{equation}
\sum_{l=1}^{3|\Gamma|} q_v(u_l) |u_l|^2 = \zeta_v, \quad (v=0, \cdots, |\Gamma|-1)
\label{eq:D}
\end{equation}
where $\zeta_v$ is the Fayet-Iliopoulos parameter satisfying $\sum_v \zeta_v=0$.
In the graph-theoretic language, the $|\Gamma| \times 3|\Gamma|$ matrix $d$
with entries $d_{v, l}=q_v(u_l)$ is the incidence matrix of the quiver,
where $d_{v, l}$ is $-1$ if $v$ is the tail of $l$, +1 if $v$ is the head of $l$ and zero otherwise.

Since the gauge group is $U(1)^{|\Gamma|-1}$ as noted above,
one of the D-flatness conditions in ($\ref{eq:D}$) is redundant.
So if we define a $(|\Gamma|-1) \times 3|\Gamma|$ matrix $\Delta$ by deleting the last row of $d$,
the $|\Gamma|-1$ independent D-flatness conditions are written as
\begin{equation}
\sum_{l=1}^{3|\Gamma|} \Delta_{v, l} |u_l|^2 =\zeta_v.
\quad (v=1, \cdots, |\Gamma|-1)
\end{equation}

To combine these conditions to the expression of $\cal Z$ ($\ref{eq:holomorphic}$),
we must  rewrite these conditions to those with respect to the coordinates $p_\lambda$,
that is, we must find the $U(1)^{|\Gamma|-1}$ charges for $p_\lambda$.
The charge matrix $Q_D$ for $p_\lambda$ is determined from the charge matrix $\Delta$ for $u_l$
and the relation between the coordinates $u_l$ and $p_\lambda$ ($\ref{eq:up}$) as
\begin{equation}
Q_D (KT)^t=\Delta.
\end{equation}

The total number of D-flatness conditions to be imposed on the 
auxiliary gauge theory is $(c-|\Gamma|-2)+(|\Gamma|-1)=c-3$, and gauge 
symmetry is $U(1)^{c-|\Gamma|-2} \times U(1)^{|\Gamma|-1}=U(1)^{c-3}$.
Thus the moduli space $\cal M$ takes the form
\begin{equation}
{\cal M}=\{(p_1,...,p_c) \in {\bf C}^c|(c-3) \; 
\mbox{D-flatness conditions}\}/U(1)^{c-3}.
\label{eq:moduli}
\end{equation}
Note that the $|\Gamma|-1$ D-flatness conditions in equation 
($\ref{eq:D}$) have Fayet-Iliopoulos parameters 
$\theta_i$, while $c-|\Gamma|-2$ D-flatness conditions coming from 
F-flatness conditions do not have such parameters.

Finally the moduli space can be obtained in the form:
\begin{equation}
{\cal M} = ({\cal Z}-S_\zeta)/({\bf C}^*)^{|\Gamma|-1}=({\bf C}^c-Z_\zeta)/({\bf C}^*)^{c-3},
\label{eq:M}
\end{equation}
where $S_\zeta, Z_\zeta$ are certain exceptional sets.
The action of $({\bf C}^*)^{c-3}$ on ${\bf C}^c$ is summarized in the $(c-3) \times c$ matrix
\begin{equation}
Q_t=\left(
\begin{array}{c}
Q_F \\
Q_D
\end{array}
\right),
\end{equation}
and the columns of the $3 \times c$ matrix $G_t={\rm Ker} (Q_t)^t$
play the role of toric generators of $\cal M$.

Before ending this section,
we show that the matrix $G_t$ is written as
\begin{equation}
G_t=\Pi K T
\end{equation}
if we introduce the $3 \times c$ matrix
\begin{equation}
\Pi=\left(
\begin{array}{ccccccccc}
1&\cdots&1&0&\cdots&0&0&\cdots&0 \\
0&\cdots&0&1&\cdots&1&0&\cdots&0 \\
0&\cdots&0&0&\cdots&0&1&\cdots&1
\end{array}
\right),
\end{equation}
where each row has $|\Gamma|$ one and $2|\Gamma|$ zero.
Since the matrix $(G_t)^t$ is defined as a kernel of $Q_t$,
what we have to show is
\begin{equation}
Q_t (\Pi KT)^t =
\left(
\begin{array}{c}
Q_F (\Pi KT)^t\\
Q_D (\Pi KT)^t
\end{array}
\right) = 0.
\end{equation}
Firstly, the equation $Q_F (\Pi KT)^t = Q_F T^t K^t \Pi^t=0$ follows from the equation $Q_F T^t=0$,
which is obtained by the definition of $Q_F$.
Secondly,
\begin{eqnarray}
&&Q_D (\Pi KT)^t = Q_D (KT)^t \Pi^t = \Delta \Pi^t \nonumber \\
&&= \left(
\begin{array}{ccc}
\sum_w \, q_1 (x_w) &
\sum_w \, q_1 (y_w) &
\sum_w \, q_1 (z_w) \\
\vdots&\vdots&\vdots \\
\sum_w \, q_{|\Gamma|-1} (x_w)
&\sum_w \, q_{|\Gamma|-1} (y_w)
&\sum_w \, q_{|\Gamma|-1} (z_w)
\end{array}
\right)=0.
\end{eqnarray}
since the $U(1)$ charge of $x_w$, for example, satisfy
\begin{equation}
\sum_{w=0}^{|\Gamma|-1} \, q_v (x_w)=\sum_{w=0}^{|\Gamma|-1} \, (\delta_{w,v+v_1}-\delta_{w, v})=0.
\end{equation}
Thus the matrix $G_t$ takes the form
\begin{equation}
G_t=\Pi KT=\left(
\begin{array}{ccc}
\sum_v {\mbox{\boldmath $m$}}(x_v) \cdot {\mbox{\boldmath $n$}}(p_1) &
\cdots &
\sum_v {\mbox{\boldmath $m$}}(x_v) \cdot {\mbox{\boldmath $n$}}(p_c) \\
\sum_v {\mbox{\boldmath $m$}}(y_v) \cdot {\mbox{\boldmath $n$}}(p_1) &
\cdots &
\sum_v {\mbox{\boldmath $m$}}(y_v) \cdot {\mbox{\boldmath $n$}}(p_c) \\
\sum_v {\mbox{\boldmath $m$}}(z_v) \cdot {\mbox{\boldmath $n$}}(p_1) &
\cdots &
\sum_v {\mbox{\boldmath $m$}}(z_v) \cdot {\mbox{\boldmath $n$}}(p_c) \label{eq:Gt}
\end{array}
\right)
\end{equation}
and the generators of the toric diagram for $\cal M$ are written as
\begin{equation}
\hat {\mbox{\boldmath $n$}}_\lambda = \left(
\begin{array}{c}
\sum_v {\mbox{\boldmath $m$}}(x_v) \cdot {\mbox{\boldmath $n$}}(p_\lambda) \\
\sum_v {\mbox{\boldmath $m$}}(y_v) \cdot {\mbox{\boldmath $n$}}(p_\lambda) \\
\sum_v {\mbox{\boldmath $m$}}(z_v) \cdot {\mbox{\boldmath $n$}}(p_\lambda) \label{eq:toric}
\end{array}
\right). \quad (\lambda=1, \cdots , c)
\end{equation}

What we would like to stress is that to obtain ${\mbox{\boldmath $n$}}_\lambda$
we only need the inner product between
${\mbox{\boldmath $m$}}(z_v)$ and ${\mbox{\boldmath $n$}}(p_\lambda)$
and the expression ${\mbox{\boldmath $n$}}_\lambda$ itself is not necessary.
Based on this observation, we propose a method to calculate the matrix $G_t$ without solving the dual cone.
Our strategy is as follows; we start with the following relation between $u$ and $p$
\begin{equation}
u_l=\prod_{\lambda=1}^c p_\lambda^{\alpha_{l \lambda}},
\end{equation}
although the number $c$ is not determined yet.
Next we impose the F-flatness condition on the indices $\alpha_{l \lambda} \in {\mbox{\boldmath $Z$}}_{\geq 0}$
and count all possible combinations of indices.
The number $c$ is determined as the number of independent combinations
which satisfies the F-flatness conditions.

\section{${\bf C}^3/{\mbox{\boldmath $Z$}}_n \times {\mbox{\boldmath $Z$}}_n$}

In this section we consider the gauge theory on the worldvolume of a D3-brane transverse to the
quotient singularity ${\bf C}^3/{\mbox{\boldmath $Z$}}_n \times {\mbox{\boldmath $Z$}}_n$ \cite{HU, Greene, MR, MT}.
We show that the calculation of  the dual cone is replaced by a tiling problem of a torus by rhombus,
and a certain dimer problem. (Similar obserbation was made in \cite{FHKVW}.)
We furthermore show that it is equivalent to a calculation of ground states of
an antiferromagnetic Ising model on a triangular lattice defined on the McKay quiver.

\subsection{Configuration of indices on the McKay quiver}

\begin{figure}[htdp]
\begin{center}
\leavevmode
\includegraphics[width=100mm,clip]{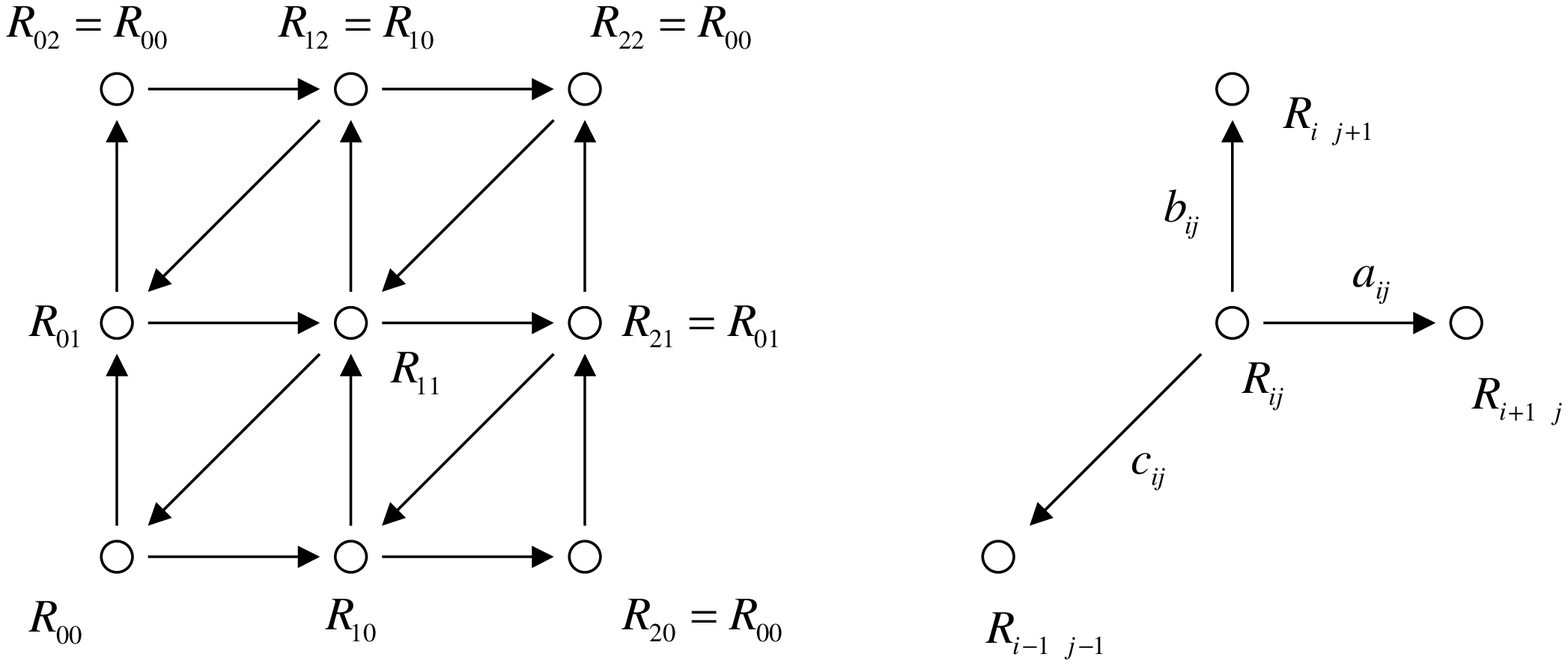}
\caption{McKay quiver of the quotient singularity ${\bf C}^3/{\mbox{\boldmath $Z$}}_2 \times {\mbox{\boldmath $Z$}}_2$}
\label{fig:McKay22}
\end{center}
\end{figure}

The group ${\mbox{\boldmath $Z$}}_n \times {\mbox{\boldmath $Z$}}_n$ has $n^2$ one dimensional irreducible representations $R_{ij}$,
and apart from the overall $U(1)$, we have a $U(1)^{n^2-1}$ gauge symmetry.
If we choose the three-dimensional representation as $R=R_{1 \, 0} \oplus R_{0 \, 1} \oplus R_{n-1 \, n-1}$,
the generators $g_1, g_2$ of the group
${\mbox{\boldmath $Z$}}_n \times {\mbox{\boldmath $Z$}}_n$ act on ${\bf C}^3$ as
\begin{eqnarray}
g_1 &:& (X^1, X^2, X^3) \rightarrow (\omega X^1, X^2, \omega^{-1} X^3), \nonumber \\
g_2 &:& (X^1, X^2, X^3) \rightarrow (X^1, \omega X^2, \omega^{-1} X^3),
\end{eqnarray}
where $\omega=e^\frac{2 \pi i}{n}$.
Then we have the following decomposition
\begin{equation}
R \otimes R_{ij}= R_{i+1 \, j} \oplus R_{i \, j+1} \oplus R_{i-1 \, j-1},
\end{equation}
and hence there are three types of arrows
\begin{eqnarray}
a_{ij} &:& R_{ij} \rightarrow R_{i+1 \, j}, \nonumber \\
b_{ij} &:& R_{ij} \rightarrow R_{i \, j+1}, \\
c_{ij} &:& R_{ij} \rightarrow R_{i-1 \, j-1}. \nonumber
\end{eqnarray}
The McKay quiver
of the quotient singularity ${\bf C}^3/{\mbox{\boldmath $Z$}}_n \times {\mbox{\boldmath $Z$}}_n$ is depicted in
Figure \ref{fig:McKay22}.

This model has $3n^2$ variables $x_{ij}, y_{ij}, z_{ij}$,
which corresponds to the arrows $a_{ij}, b_{ij}, c_{ij}$ respectively.
The F-flatness conditions on these variables take the form
\begin{eqnarray}
&& x_{ij} \, y_{i+1 \, j} = y_{ij} \, x_{i \, j+1}, \nonumber \\
&& y_{ij} \, z_{i \, j+1} = z_{ij} \, y_{i-1 \, j-1}, \label{eq:Fnn}\\
&& z_{ij} \, x_{i-1 \, j-1} = x_{ij} \, z_{i+1 \, j}. \nonumber
\end{eqnarray}
As explained in the last section, we introduce new variables $p_\lambda$ ($\lambda=1, \cdots, c$) which are related to
the original variables $x_{ij}, y_{ij}, z_{ij}$ ($i,j=0, \cdots, n-1$)
by the following equations,
\begin{eqnarray}
x_{ij}&=&p_1^{\alpha_{ij}^1} \; p_2^{\alpha_{ij}^2} \; \cdots \; p_c^{\alpha_{ij}^c}, \nonumber \\
y_{ij}&=&p_1^{\beta_{ij}^1} \; p_2^{\beta_{ij}^2} \; \cdots \; p_c^{\beta_{ij}^c}, \label{eq:xyzp} \\
z_{ij}&=&p_1^{\gamma_{ij}^1} \; p_2^{\gamma_{ij}^2} \; \cdots \; p_c^{\gamma_{ij}^c}, \nonumber
\end{eqnarray}
where the indices $\alpha_{ij}^\lambda, \beta_{ij}^\lambda, \gamma_{ij}^\lambda \in {\mbox{\boldmath $Z$}}_{\geq 0}$
are subject to the F-flatness conditions.
If we consider the quantity $x_{ij} \, y_{i+1 \, j} \, z_{i+1 \, j+1}$,
we obtain the following relations
\begin{eqnarray}
x_{ij} \, y_{i+1 \, j} \, z_{i+1 \, j+1}
&=& y_{ij} \, x_{i \, j+1} \, z_{i+1 \, j+1} \nonumber \\
&=& y_{ij} \, z_{i \, j+1} \, x_{i-1 \, j}  \label{eq:XYZ} \\
&=& z_{ij} \, y_{i-1 \, j-1} \, x_{i-1 \, j}, \nonumber
\end{eqnarray}
due to the F-flatness conditions (\ref{eq:Fnn}).
In terms of the indices $\alpha_{ij}^\lambda, \beta_{ij}^\lambda, \gamma_{ij}^\lambda$,
these relations are written as
\begin{eqnarray}
\alpha_{ij}^\lambda + \beta_{i+1 \, j}^\lambda + \gamma_{i+1 \, j+1}^\lambda
&=& \beta_{ij}^\lambda + \alpha_{i \, j+1}^\lambda + \gamma_{i+1 \, j+1}^\lambda \nonumber \\
&=& \beta_{ij}^\lambda + \gamma_{i \, j+1}^\lambda + \alpha_{i-1 \, j}^\lambda \label{eq:triangle} \\
&=& \gamma_{ij}^\lambda + \beta_{i-1 \, j-1}^\lambda + \alpha_{i-1 \, j}^\lambda. \nonumber
\end{eqnarray}

\begin{figure}[htdp]
\begin{center}
\leavevmode
\includegraphics[width=50mm,clip]{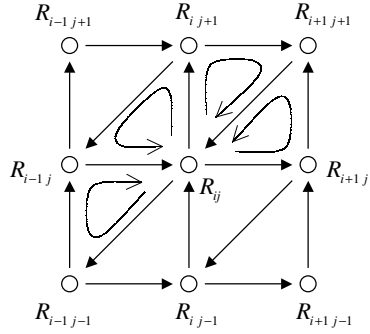}
\caption{F-flatness conditions imply the equivalence among triangles in the McKay quiver.}
\label{fig:triangle}
\end{center}
\end{figure}

To investigate sets of indices $\alpha_{ij}^\lambda, \beta_{ij}^\lambda, \gamma_{ij}^\lambda$
satisfying these conditions, it is useful to consider configurations of indices on the McKay quiver.
As noted above,
each variables $x_{ij}, \, y_{i+1 \, j}$ and $z_{i+1 \, j+1}$
corresponds to arrows $a_{ij}$, $b_{i+1 \, j}$ and $c_{i+1 \, j+1}$.
Thus in terms of the arrows, the relations (\ref{eq:XYZ}) are written as
\begin{eqnarray}
a_{ij} \, b_{i+1 \, j} \, c_{i+1 \, j+1}
&=& b_{ij} \, a_{i \, j+1} \, c_{i+1 \, j+1} \nonumber \\
&=& b_{ij} \, c_{i \, j+1} \, a_{i-1 \, j} \\
&=& c_{ij} \, b_{i-1 \, j-1} \, a_{i-1 \, j}. \nonumber
\end{eqnarray}
These relations imply that all closed paths encircling triangles of the McKay quiver are equivalent
as depicted in Figure \ref{fig:triangle}.

\begin{figure}[htdp]
\begin{center}
\leavevmode
\includegraphics[width=60mm,clip]{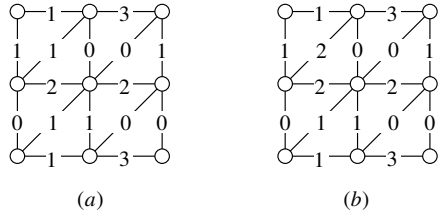}
\caption{Configurations of indices on the McKay quiver for ${\bf C}^3/{\mbox{\boldmath $Z$}}_2 \times {\mbox{\boldmath $Z$}}_2$.
$({\rm a})$ is a configuration with $s=3$.
$(b)$ is not allowed since sums of indices are different from triangle to triangle.}
\label{fig:sum}
\end{center}
\end{figure}

If we assign the indices $\alpha_{ij}^\lambda$, $\beta_{ij}^\lambda$ and $\gamma_{ij}^\lambda$
to arrows $a_{ij}$, $b_{i+1 \, j}$ and $c_{i+1 \, j+1}$ respectively,
we obtain a configuration of indices on the McKay quiver.
The conditions ($\ref{eq:triangle}$) imply that the sum of indices on every triangle of the McKay quiver must be equal.
Thus we only need to consider configurations of indices satisfying this condition.
We denote the sum of the indices of each triangle by $s$ (See Figure \ref{fig:sum}).

\begin{figure}[htdp]
\begin{center}
\leavevmode
\includegraphics[width=120mm,clip]{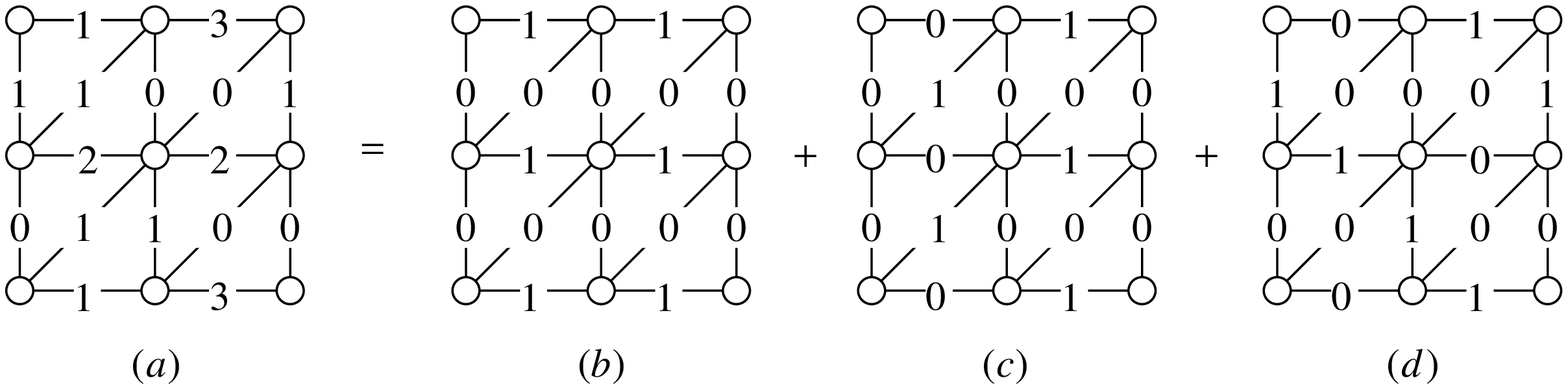}
\caption{A configuration with $s=3$ is written as a
superposition of three configurations with $s=1$.}
\label{fig:superposition}
\end{center}
\end{figure}

Here an important point is that any configuration with the sum $s$ can be written
as a "superposition" of $s$ configurations with $s=1$ (See Figure \ref{fig:superposition}).
Based on this fact we assert that
each configuration with $s=1$ corresponds to a coordinate $p_\lambda$ 
and hence $c$ is the number ofdifferent configurations with $s=1$.
Thus what we have to do is to enumerate all configurations of indices with $s=1$.

\begin{figure}[htdp]
\begin{center}
\leavevmode
\includegraphics[width=160mm,clip]{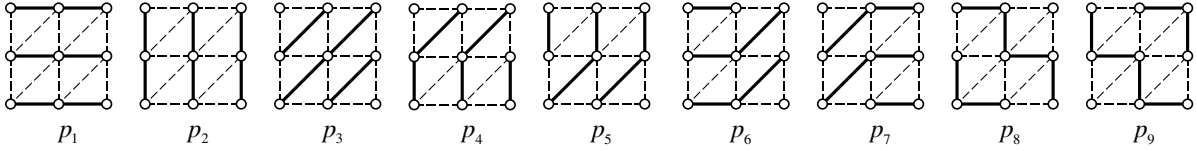}
\caption{Configurations with $s=1$ on the McKay quiver for ${\bf C}^3/{\mbox{\boldmath $Z$}}_2 \times {\mbox{\boldmath $Z$}}_2$.
Bold lines represent links assigned index one and thin lines represent links assigned index zero.
Each configuration corresponds to a coordinate $p_\lambda$.}
\label{fig:index22}
\end{center}
\end{figure}

As an example we depict all configurations of indices with $s=1$ for ${\bf C}^3/{\mbox{\boldmath $Z$}}_2 \times {\mbox{\boldmath $Z$}}_2$
in Figure \ref{fig:index22}.

There are nine configurations with $s=1$, which means that $c=9$.
By using the equation (\ref{eq:xyzp}), we obtain the relation between $x_{ij}, y_{ij}, z_{ij}$ and $p_1, \cdots, p_9$ as follows;
\begin{eqnarray}
&&x_{00}=p_1 p_6 p_8, \quad
x_{10}=p_1 p_7 p_9, \quad
x_{01}=p_1 p_6 p_9, \quad
x_{11}=p_1 p_7 p_8, \nonumber \\
&&y_{00}=p_2 p_4 p_8, \quad
y_{10}=p_2 p_4 p_9, \quad
y_{01}=p_2 p_5 p_9, \quad
y_{11}=p_2 p_5 p_8, \\
&&z_{00}=p_3 p_4 p_6, \quad
z_{10}=p_3 p_4 p_7, \quad
z_{01}=p_3 p_5 p_6, \quad
z_{11}=p_3 p_5 p_7. \nonumber
\end{eqnarray}
From the equation (\ref{eq:up}), we obtain a matrix $KT$ as
\begin{equation}
K T
=(\mbox{\boldmath $m$}_{l} \cdot \mbox{\boldmath $n$}_\lambda) 
=\left(
\begin{array}{cccccccccc}
&\mbox{\boldmath $n$}_1&\mbox{\boldmath $n$}_2&\mbox{\boldmath $n$}_3&
\mbox{\boldmath $n$}_4&\mbox{\boldmath $n$}_5&\mbox{\boldmath $n$}_6&
\mbox{\boldmath $n$}_7&\mbox{\boldmath $n$}_8&\mbox{\boldmath $n$}_9\\
\mbox{\boldmath $m$}_{1} \;(x_{00})&1&0&0&0&0&1&0&1&0\\
\mbox{\boldmath $m$}_{2} \;(x_{10})&1&0&0&0&0&0&1&0&1\\
\mbox{\boldmath $m$}_{3} \;(x_{01})&1&0&0&0&0&1&0&0&1\\
\mbox{\boldmath $m$}_{4} \;(x_{11})&1&0&0&0&0&0&1&1&0\\
\mbox{\boldmath $m$}_{5} \;(y_{00})&0&1&0&1&0&0&0&1&0\\
\mbox{\boldmath $m$}_{6} \;(y_{10})&0&1&0&1&0&0&0&0&1\\
\mbox{\boldmath $m$}_{7} \;(y_{01})&0&1&0&0&1&0&0&0&1\\
\mbox{\boldmath $m$}_{8} \;(y_{11})&0&1&0&0&1&0&0&1&0\\
\mbox{\boldmath $m$}_{9} \;(z_{00})&0&0&1&1&0&1&0&0&0\\
\mbox{\boldmath $m$}_{10} \;(z_{10})&0&0&1&1&0&0&1&0&0\\
\mbox{\boldmath $m$}_{11} \;(z_{01})&0&0&1&0&1&1&0&0&0\\
\mbox{\boldmath $m$}_{12} \;(z_{11})&0&0&1&0&1&0&1&0&0
\end{array}
\right).
\label{eq:nm}
\end{equation}
This result is the same as that obtained by the calculation of dual cone.

\begin{figure}[htdp]
\begin{center}
\leavevmode
\includegraphics[width=35mm,clip]{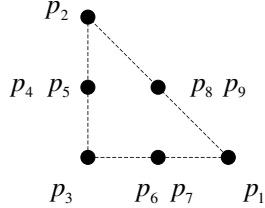}
\caption{Toric diagram for the model ${\bf C}^3/{\mbox{\boldmath $Z$}}_2 \times {\mbox{\boldmath $Z$}}_2$.
We assign a coordinate $p_\lambda$ to each toric generator.}
\label{fig:degeneracy22}
\end{center}
\end{figure}

By defining the charge projection matrix $\Pi$ as
\begin{equation}
\Pi = \left(
\begin{array}{cccccccccccc}
1&1&1&1&0&0&0&0&0&0&0&0\\
0&0&0&0&1&1&1&0&0&0&0&0\\
0&0&0&0&0&0&0&0&1&1&1&1
\end{array}
\right),
\label{eq:projection}
\end{equation}
we obtain nine toric generators $\hat {\mbox{\boldmath $n$}}_\lambda$
according to the equation (\ref{eq:Gt}), 
\begin{equation}
G_t=\Pi K T=\left(
\begin{array}{ccccccccc}
\hat {\mbox{\boldmath $n$}}_1&\hat {\mbox{\boldmath $n$}}_2&\hat {\mbox{\boldmath $n$}}_3&
\hat {\mbox{\boldmath $n$}}_4&\hat {\mbox{\boldmath $n$}}_5&\hat {\mbox{\boldmath $n$}}_6&
\hat {\mbox{\boldmath $n$}}_7&\hat {\mbox{\boldmath $n$}}_8&\hat {\mbox{\boldmath $n$}}_9\\
4&0&0&0&0&2&2&2&2\\
0&4&0&2&2&0&0&2&2\\
0&0&4&2&2&2&2&0&0
\end{array}
\right).
\end{equation}
Thus we finally obtain the toric diagram as depicted in Figure \ref{fig:degeneracy22}.

\subsection{Enumeration of configuration of indices on the McKay quiver}

Now we would like to present a systematic way to count the configurations of indices with $s=1$.
As we will see below, the configuration of indices is translated to other systems such as a tiling problem, a dimer problem and an Ising model.

\begin{figure}[htdp]
\begin{center}
\leavevmode
\includegraphics[width=120mm,clip]{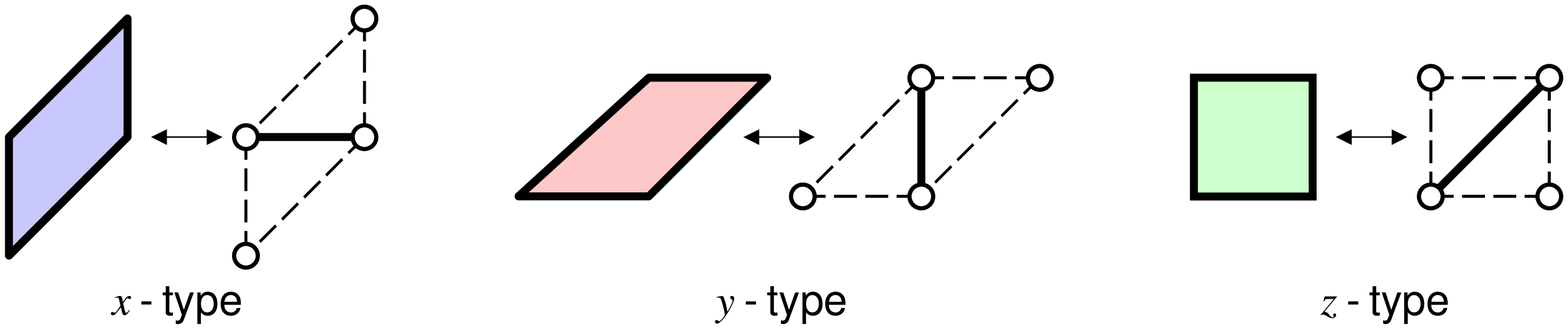}
\caption{Three types of rhombuses.
Each rhombus should be put on the McKay quiver such that its diagonal lies on the indices with one.}
\label{fig:rhombus}
\end{center}
\end{figure}

Firstly, we show that enumerating configurations of indices with $s=1$
on the McKay quiver for ${\bf C}^3/{\mbox{\boldmath $Z$}}_n \times {\mbox{\boldmath $Z$}}_n$
is equivalent to a certain tiling problem.
What is tiled is a torus with size $n \times n$
and tiles are three types of rhombuses depicted in Figure \ref{fig:rhombus}.

\begin{figure}[htdp]
\begin{center}
\leavevmode
\includegraphics[width=160mm,clip]{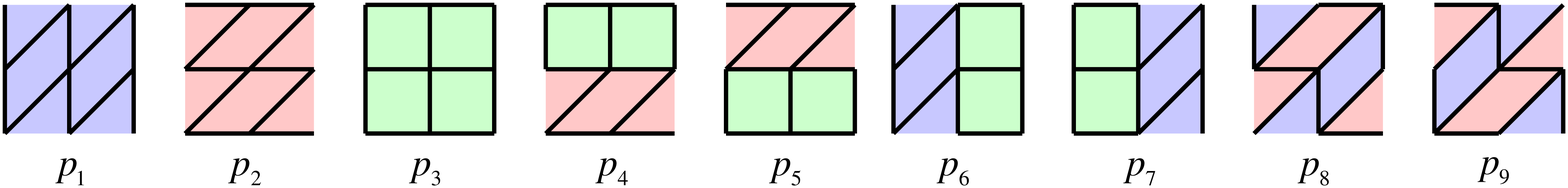}
\caption{Tiling a torus with size $2 \times 2$ by rhombus.}
\label{fig:tiling22}
\end{center}
\end{figure}

Each rhombus should be put on the McKay quiver such that its diagonal lies on the link with indices one
and its boundary on the link with indices zero.
Then the conditions (\ref{eq:triangle}) on the indices assures that rhombuses cover the torus
and do not overlap each other.
The fact that configurations of indices with the sum $s$ is a superposition of $s$ configurations with $s=1$
corresponds to the fact that a $s$-tiple tiling is decomposed to $s$ tilings with $s=1$.
As an example, we depict all tilings for the model ${\bf C}^3/{\mbox{\boldmath $Z$}}_2 \times {\mbox{\boldmath $Z$}}_2$
in Figure \ref{fig:tiling22}.

Secondly, the configuration of indices is translated to a dimer problem
on a hexagonal lattice.
The vertices of the lattice lie on the center of the triangles of the McKay quiver
and dimers connect neighboring vertices such  that they intersect to links with indices one.

Thirdly, the problem to enumerate configurations of indices with $s=1$ is equivalent to a problem
to count ground states of an Ising model on a lattice defined on the quiver diagram.
Lattice sites of this model are the nodes $R_{ij}$ of the McKay quiver,
and spin variables $\pm$ at the nodes interact through neighboring links connected by arrows.
Thus the lattice is a triangular one with $n \times n$ sites on a torus.
We consider an antiferromagnetic Ising model on this lattice.
Although an antiferromagnetic Ising model favors anti-parallel links ($+-$) and ($-+$),
there must exist parallel links ($++$) and ($--$) due to frustration since the lattice is triangular.
Therefore every triangle in the lattice has two parallel links and one anti-parallel link in the ground states.
By deleting parallel links from the lattice, we obtain a diagram of anti-parallel links.
Anti-parallel links form rhombuses because of the property of the ground state of the Ising model.
Thus each ground state of the Ising model corresponds to a tiling by rhombuses discussed above.

Note that not only periodic boundary condition but also ant-periodic conditions are allowed.
Thus there are four boundary conditions
\begin{eqnarray}
({\rm p},{\rm p})&&\sigma_{ij}=\sigma_{i+n \, j}, \quad \sigma_{ij}=\sigma_{i \, j+n}, \nonumber \\
({\rm p},{\rm a})&&\sigma_{ij}=\sigma_{i+n \, j}, \quad \sigma_{ij}=-\sigma_{i \, j+n}, \nonumber \\
({\rm a},{\rm p})&&\sigma_{ij}=-\sigma_{i+n \, j}, \quad \sigma_{ij}=\sigma_{i \, j+n}, \\
({\rm a},{\rm a})&&\sigma_{ij}=-\sigma_{i+n \, j}, \quad \sigma_{ij}=-\sigma_{i \, j+n}. \nonumber
\end{eqnarray}
where $"{\rm p}"$ and $"{\rm a}"$ represent periodic and anti-periodic boundary conditions.
An important point is that interchange of $+$ spin and $-$ spin gives the identical configuration,
the correspondence between ground states of the Ising model and tilings is two to one.

\begin{figure}[htdp]
\begin{center}
\leavevmode
\includegraphics[width=160mm,clip]{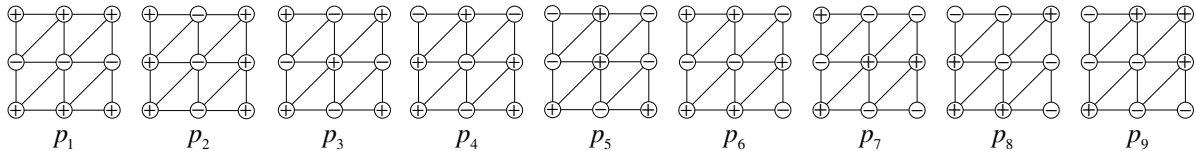}
\caption{Ground states with $\sigma_{00}=+$ of the antiferromagnetic Ising model
on the lattice of the McKay quiver for ${\bf C}^3/{\mbox{\boldmath $Z$}}_2 \times {\mbox{\boldmath $Z$}}_2$.
}
\label{fig:Ising22}
\end{center}
\end{figure}

Ground states of the Ising model on the McKay quiver for
${\bf C}^3/{\mbox{\boldmath $Z$}}_2 \times {\mbox{\boldmath $Z$}}_2$ is depicted in Figure \ref{fig:Ising22}.
The boundary conditions for $\{p_1, p_2, p_3\}$, $\{p_4, p_5\}$, $\{p_6, p_7\}$ and $\{p_8, p_9\}$
are $({\rm p},{\rm p})$, $({\rm p},{\rm a})$, $({\rm a},{\rm p})$ and $({\rm a},{\rm a})$, respectively.

\section{Counting the number of the ground states of the Ising model}
\reseteqnum

In this section we count the number of ground states of the Ising model
introduced in the last section.

The energy $E$ of the Ising model on the triangular lattice is given by
\begin{equation}
E=J \sum_{i,j=0}^{n-1} (\sigma_{i j} \sigma_{i+1 j} +\sigma_{i j} \sigma_{i j+1} +\sigma_{i+1 j+1} \sigma_{i j}).
\end{equation}
For an antiferromagnetic Ising model, the coupling constant $J$ is positive.
The partition function of the Ising model is
\begin{equation}
Z_{n \times n} = \sum_{\{\sigma_{ij}\}} {\rm e}^{-\beta E},
\end{equation}
where the sum is taken over all configurations of spins $\{\sigma_{ij}\}$
which satisfy (anti-) periodic boundary conditions
in the horizontal and vertical directions.
Therefore the partition function is decomposed into a sum of partition functions with the four boundary conditions,
\begin{equation}
Z_{n \times n} = Z_{n \times n}^{({\rm pp})} +Z_{n \times n}^{({\rm pa})} +Z_{n \times n}^{({\rm ap})} + Z_{n \times n}^{({\rm aa})}.
\end{equation}
In $Z_{n \times n}^{({\rm pp})}$, for example, summation is restricted to configurations of spins $\{\sigma_{ij}\}$
with the (p, p) boundary condition.

To calculate $Z_{n \times n}^{({\rm pp})}$, we introduce a transfer matrix $T_n^{({\rm p})}$
as follows;
\begin{eqnarray}
Z_{n \times n}^{({\rm pp})}
&=& \sum_{\{\sigma_{ij}\}} \prod_{j=0}^{n-1} \prod_{i=0}^{n-1}
{\rm e}^{-\beta J \left[
\frac{1}{2} (\sigma_{i j} \sigma_{i+1 j} +\sigma_{i j+1} \sigma_{i+1 j+1})
+\sigma_{i j} \sigma_{i j+1}
+\sigma_{i+1 j+1} \sigma_{i j}
\right]} \nonumber\\
&=& \sum_{\{\sigma_{ij}\}} \prod_{j=0}^{n-1} (T_n^{({\rm p})})_{\sigma_j \sigma_{j+1}} \\
&=& {\rm Tr} \, (T_n^{({\rm p})})^n. \nonumber
\end{eqnarray}
The suffix $\sigma_j$ of the transfer matrix $T_n^{({\rm p})}$ represents a set of spin variables of the
$j$-th row $\{\sigma_{0 j}, \cdots , \sigma_{n-1 j}\}$,
so $T_n^{({\rm p})}$ has $2^n \times 2^n$ elements.
We use the lexicographic order for the suffix of the matrix.
Using the transfer matrix $T_n^{({\rm p})}$,
the partition function for the (pa) boundary condition
is written as
\begin{equation}
Z_{n \times n}^{({\rm pa})}
= {\rm Tr} \, [(T_n^{({\rm p})})^n R_n],
\end{equation}
where the $2^n \times 2^n$ matrix
\begin{equation}
R_n=\left(
\begin{array}{cc}
0&1\\
1&0
\end{array}
\right)^{\otimes n}
\end{equation}
is inserted in order to realize the anti-periodic boundary conditions in the vertical direction:
\begin{equation}
\sigma_{j n}=-\sigma_{j 0}.
\end{equation}
Similarly, the partition functions for the boundary conditions (ap) and (aa)
are written as
\begin{eqnarray}
Z_{n \times n}^{({\rm ap})} &=& {\rm Tr} (T_n^{({\rm a})})^n, \\
Z_{n \times n}^{({\rm aa})} &=& {\rm Tr} [(T_n^{({\rm a})})^n R_n],
\end{eqnarray}
where $T_n^{({\rm a})}$ is the transfer matrix for the anti-periodic boundary condition in the horizontal direction:
\begin{equation}
\sigma_{n j}=-\sigma_{0 j}.
\end{equation}

\begin{figure}[htdp]
\begin{center}
\leavevmode
\includegraphics[width=100mm,clip]{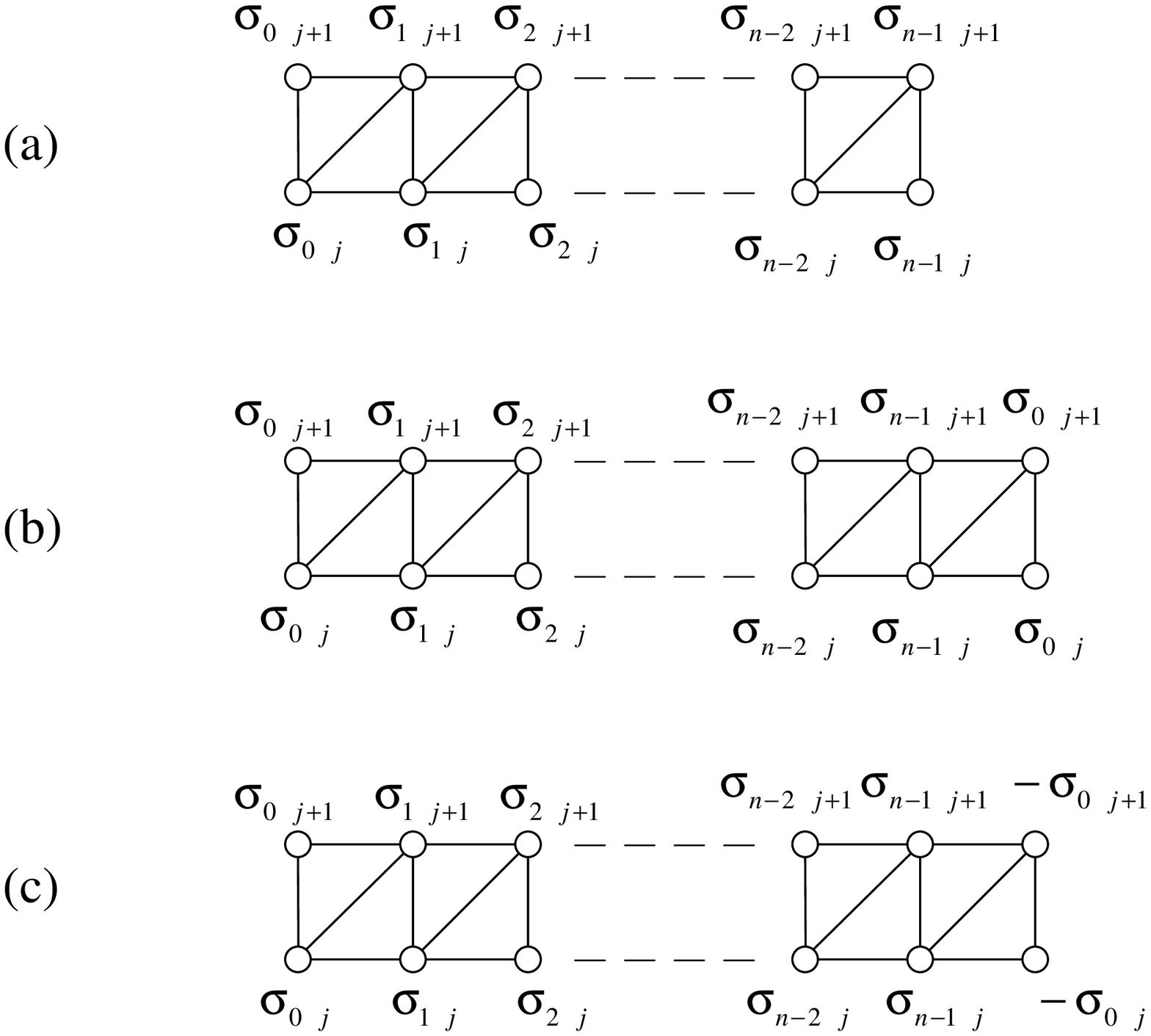}
\caption{lattice}
\label{fig:latticen}
\end{center}
\end{figure}

What we have to do next is to find the transfer matrices $T_n^{({\rm p})}$ and $T_n^{({\rm a})}$.
For this purpose,
we consider configurations of spins on the $2 \times n$ lattice depicted in Figure \ref{fig:latticen} (a)
with free boundary conditions.

The $2^n \times 2^n$ transfer matrix $T_n$ corresponding to this configuration is written as
\begin{equation}
(T_n)_{\sigma_j \sigma_{j+1}}
=\prod_{i=0}^{n-2} {\rm e}^{-\beta J \left[
\frac{1}{2} (\sigma_{i j} \sigma_{i+1 j} +\sigma_{i j+1} \sigma_{i+1 j+1})
+\sigma_{i j} \sigma_{i j+1}
+\sigma_{i+1 j+1} \sigma_{i j}
\right]}. \nonumber\\
\end{equation}
This matrix satisfies a following recursion relation
\begin{equation}
(T_n)_{\sigma \sigma'} = (T_{n-1} \otimes H_1)_{\sigma \sigma'} (H_{n-2} \otimes T_2)_{\sigma \sigma'}
\end{equation}
where the $2^k \times 2^k$ matrix $H_k$ is given by
\begin{equation}
H_k=\left(
\begin{array}{cc}
1&1\\
1&1
\end{array}
\right)^{\otimes \, k}.
\end{equation}
By using the explicit form of the matrix $T_2$:
\begin{equation}
T_2=\left(
\begin{array}{cccc}
{\rm e}^{-3K}&1&1&{\rm e}^{K}\\
{\rm e}^{-2K}&{\rm e}^{K}&{\rm e}^{K}&{\rm e}^{2K}\\
{\rm e}^{2K}&{\rm e}^{K}&{\rm e}^{K}&{\rm e}^{-2K}\\
{\rm e}^{K}&1&1&{\rm e}^{-3K}
\end{array}
\right)
\end{equation}
with $K=\beta J$, we are able to obtain the matrix $T_n$.
\begin{figure}[htdp]
\begin{center}
\leavevmode
\includegraphics[width=70mm,clip]{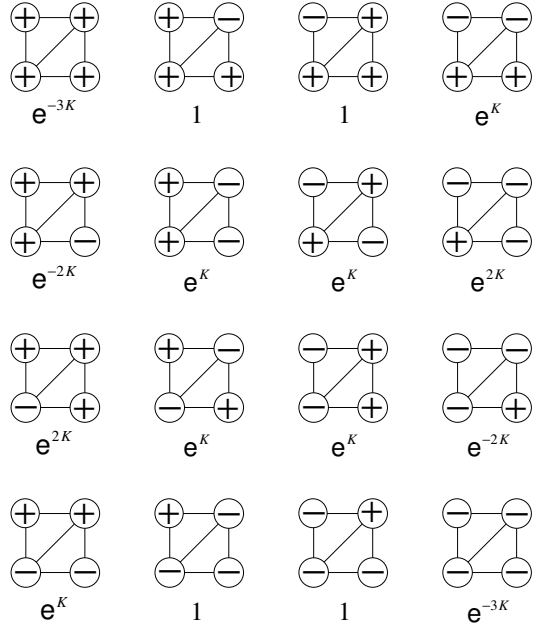}
\caption{The matrix elements of $T_2$.}
\label{fig:T2}
\end{center}
\end{figure}

Configurations of spins with the periodic boundary condition in the horizontal direction
are realized by adding a pair of spins $\sigma_{00}$ and $\sigma_{01}$
on the right side of the lattice as depicted in Figure \ref{fig:latticen} (b).
Thus the transfer matrix $T_n^{({\rm p})}$ is obtained from $T_n$ by adding the contribution from the last 
plaquet in Figure \ref{fig:latticen} (b).
Therefore we obtain
\begin{equation}
(T^{({\rm p})}_n)_{\sigma_j \sigma_{j+1}} = (T_n)_{\sigma_j \sigma_{j+1}} (P_n)_{\sigma_j \sigma_{j+1}}
\end{equation}
with
\begin{equation}
(P_n)_{\sigma_j \sigma_{j+1}}
={\rm e}^{-\beta J \left[
\frac{1}{2} (\sigma_{n-2 \, j} \sigma_{0 \, j} +\sigma_{n-2 \, j+1} \sigma_{0 \, j+1})
+\sigma_{n-2 \, j} \sigma_{n-2 \, j+1}
+\sigma_{0 \, j+1} \sigma_{n-2 \, j}
\right]}.
\end{equation}
By using the matrix $T_2$, the $2^n \times 2^n$ matrix $P_n$ is written as
\begin{equation}
P_n=X_n^t (T_2 \otimes H_{n-2}) X_n
\end{equation}
with
\begin{equation}
X_n=(X_2 \otimes {\bf 1}_2^{n-2}) ({\bf 1}_2 \otimes X_{n-1}), \qquad
X_2=\left(
\begin{array}{cccc}
1&0&0&0\\
0&0&1&0\\
0&1&0&0\\
0&0&0&1
\end{array}
\right).
\end{equation}
For example, the matrix $T_2^{({\rm p})}$ is given by
\begin{equation}
T_2^{({\rm p})}=\left(
\begin{array}{cccc}
{\rm e}^{-6k}&1&1&{\rm e}^{2K}\\
1&{\rm e}^{2K}&{\rm e}^{2K}&1\\
1&{\rm e}^{2K}&{\rm e}^{2K}&1\\
{\rm e}^{2k}&1&1&{\rm e}^{-6K}
\end{array}
\right).
\end{equation}

The transfer matrix $T_n^{({\rm a})}$ for the antiperiodic boundary condition along the horizontal direction
is obtained from $T_n$ by adding the contribution from the last 
plaquet in Figure \ref{fig:latticen} (c).
Thus $T_n^{({\rm a})}$ is written as
\begin{equation}
(T^{({\rm a})}_n)_{\sigma_j \sigma_{j+1}} = (T_n)_{\sigma_j \sigma_{j+1}} (A_n)_{\sigma_j \sigma_{j+1}}
\end{equation}
with
\begin{equation}
(A_n)_{\sigma_j \sigma_{j+1}}
={\rm e}^{-\beta J \left[
\frac{1}{2} (-\sigma_{n-2 \, j} \sigma_{0 \, j} -\sigma_{n-2 \, j+1} \sigma_{0 \, j+1})
+\sigma_{n-2 \, j} \sigma_{n-2 \, j+1}
-\sigma_{0 \, j+1} \sigma_{n-2 \, j}
\right]}.
\end{equation}
The $2^n \times 2^n$ matrix $A_n$ is given by
\begin{equation}
A_n=Z_n^t P_n Z_n
\end{equation}
where
\begin{equation}
Z_n=\left(
\begin{array}{cc}
0&1_{2^{n-1}}\\
1_{2^{n-1}}&0\\
\end{array}
\right)
\end{equation}
makes the boundary condition antiperiodic in the vertical direction.
For example, the matrix $T_2^{({\rm a})}$ is given by
\begin{equation}
T_2^{({\rm a})}=\left(
\begin{array}{cccc}
{\rm e}^{-2k}&{\rm e}^{2K}&{\rm e}^{-2K}&{\rm e}^{2K}\\
{\rm e}^{-2K}&{\rm e}^{-2K}&{\rm e}^{2K}&{\rm e}^{2K}\\
{\rm e}^{2K}&{\rm e}^{2K}&{\rm e}^{-2K}&{\rm e}^{-2K}\\
{\rm e}^{2k}&{\rm e}^{-2K}&{\rm e}^{2K}&{\rm e}^{-2K}
\end{array}
\right).
\end{equation}

Combining these results, the partition function is written as
\begin{equation}
Z_{n \times n} = {\rm Tr} [\{ (T_n^{({\rm p})})^n+(T_n^{({\rm a})})^n \} (1+R_n)].
\end{equation}
For example, we obtain
\begin{equation}
Z_{2 \times 2} = 18{\rm e}^{4K} +32 +12{\rm e}^{-4K} +2{\rm e}^{-12K}.
\end{equation}
The first term proportional to ${\rm e}^{4K}$ corresponds to the ground states.
The coefficient 18 means that the number of ground states of the Ising model is 18.
Due to the two to one correspondence noted in the last section, the number $c$ of the toric generators is 9.

\begin{figure}[htdp]
\begin{center}
\leavevmode
\includegraphics[width=80mm,clip]{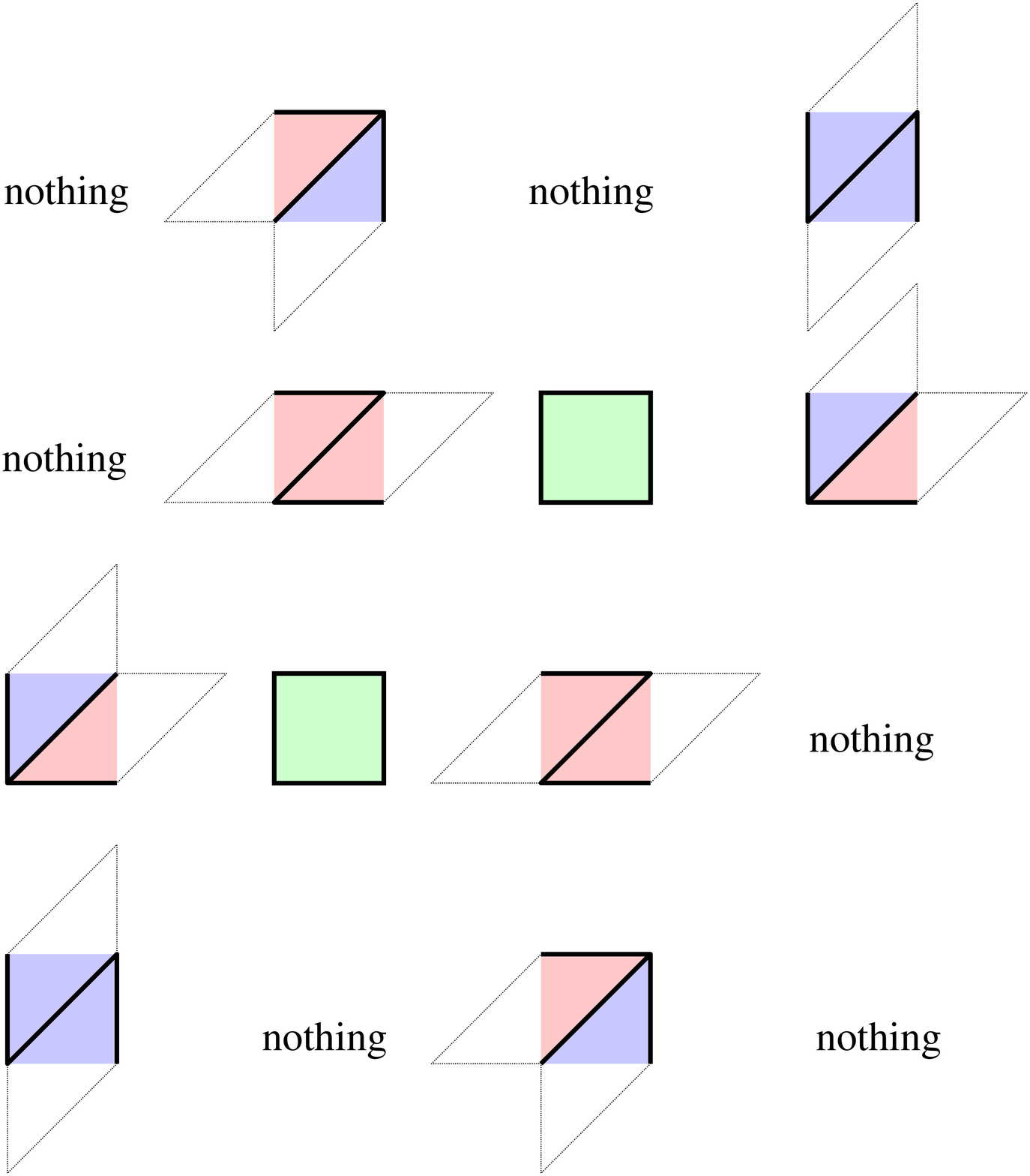}
\caption{Combinations of rhombuses corresponding to each plaquet in Figure \ref{fig:T2}.}
\label{fig:T2'}
\end{center}
\end{figure}

So far we have calculated the number $c$ of the toric generators by computing the partition function of the Ising model.
In the language of the tiling problem,
$c$ is the number of tilings of the torus by three types of rhombi.
In fact, however, this is not the only information we can extract from the calcuration of the partition function;
we are able to count the numbers of $x$- $y$- and $z$- type rhombuses contained in each tiling.
For this purpose,
we consider which rhombus corresponds to each plaquet in Figure \ref{fig:T2}.
For example, the top left plaquet (1-1 element) in Figure \ref{fig:T2}
does not corresponds to any combination of rhombuses since the three spins of each triangle have the same spins,
while the plaquet right to this one (1-2 element) corresponds to a combination
of the upper half of the $x$-type rhombus and the right half of the $y$-type rhombus.
In this way, we obtain combinations of rhombuses as depicted in Figure \ref{fig:T2'}.

\begin{figure}[htdp]
\begin{center}
\leavevmode
\includegraphics[width=130mm,clip]{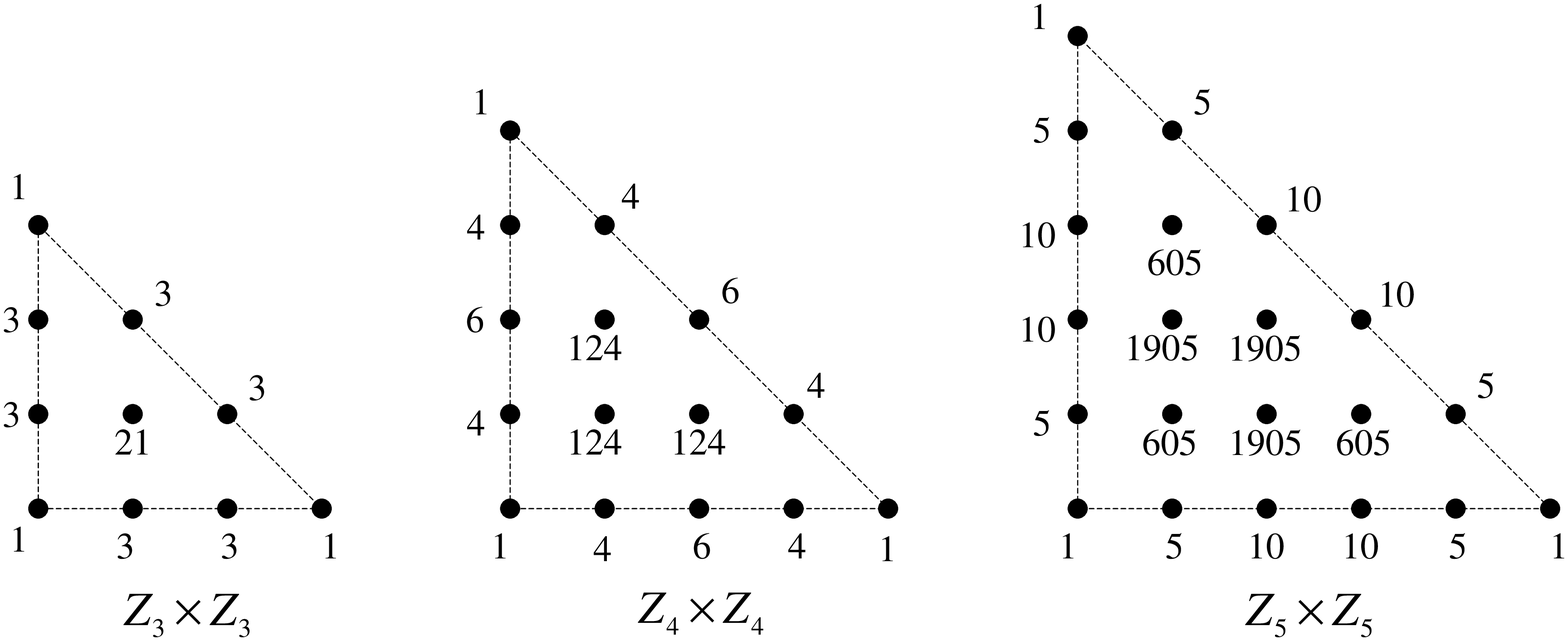}
\caption{The numbers of toric generators on each vertex.}
\label{fig:degeneracy}
\end{center}
\end{figure}

Now we assign weights that count the number of rhombuses to each element in Figure \ref{fig:T2'}.
For example, we assign zero to the top left element and $\sqrt {xy}$ to the right to this one.
These data are summarized in in a matrix as:
\begin{equation}
\hat T_2=\left(
\begin{array}{cccc}
0&\sqrt {x y}&0&\sqrt {x x}\\
0&\sqrt {y y}&z&\sqrt {x y}\\
\sqrt {x y}&z&\sqrt {y y}&0\\
\sqrt {x x}&0&\sqrt {x y}&0
\end{array}
\right).
\end{equation}
Now we introduce a function $\hat Z_{n \times n}$
obtained by replacing $T_2$ in $Z_{n \times n}$ with $\hat T_2$.
Then, for example, we obtain
\begin{equation}
\frac{1}{2} \hat Z_{2 \times 2} =x^4+y^4+z^4+2y^2 z^2+2z^2 x^2+2 x^2 y^2 + \cdots
\end{equation}
where we have omitted terms which do not correspond to the ground states.
The first term $x^4$, for example, implies that there is a tiling with four $x$-type rhombuses,
and the last term $2 x^2 y^2$ implies that there are two tilings with two $x$-type rhombuses and two $y$-type rhombuses.
Note that the numbers of  $x$- $y$- and $z$- type rhombuses
correspond to the three entries of the toric generators $\hat {\mbox{\boldmath $n$}}_\lambda$ in the equation (\ref{eq:Gt}),
and hence we are able to draw toric diagrams.
In Figure \ref{fig:degeneracy}, the numbers of toric generators on each vertex are depicted for $n=3, 4, 5$.

So far we have considered only the models of the form ${\bf C}^3/{\mbox{\boldmath $Z$}}_n \times {\mbox{\boldmath $Z$}}_n$,
but it is straightforward to generalize the analysis to the models of the form
${\bf C}^3/{\mbox{\boldmath $Z$}}_n \times {\mbox{\boldmath $Z$}}_m$
The partition function for the model $Z_{n \times m}$ is given by 
\begin{equation}
Z_{n \times m} = Z_{n \times m}^{({\rm pp})} +Z_{n \times m}^{({\rm pa})} +Z_{n \times m}^{({\rm ap})} + Z_{n \times m}^{({\rm aa})}
\end{equation}
where
\begin{eqnarray}
Z_{n \times m}^{({\rm pp})} &=& {\rm Tr} (T_n^{({\rm p})})^m, \\
Z_{n \times m}^{({\rm pa})} &=& {\rm Tr} [(T_n^{({\rm p})})^m R_n], \\
Z_{n \times m}^{({\rm ap})} &=& {\rm Tr} (T_n^{({\rm a})})^m, \\
Z_{n \times m}^{({\rm aa})} &=& {\rm Tr} [(T_n^{({\rm a})})^m R_n].
\end{eqnarray}
The calculation of the numbers of rhombuses of each type is also straightforward.
Table \ref{table:multiplicity}
shows the number $c$ of homogeneous coordinates for the model ${\bf C}^3/{\mbox{\boldmath $Z$}}_n \times {\mbox{\boldmath $Z$}}_m$
for several $n$ and $m$.
\begin{table}[htdp]
\begin{center}
\begin{tabular}{c|cccccc}
$$&1&2&3&4&5&6 \\ \hline
$1$&3&5&9&17&33&65 \\
$2$&5&9&17&33&65&129 \\
$3$&9&17&42&113&309&860 \\
$4$&17&33&113&417&1537&5793 \\
$5$&33&65&309&1537&7623&38405 \\
$6$&65&129&860&5793&38405&263640 \\
$7$&129&257&2445&22081&193849&1849073 \\
$8$&257&513&7073&84609&984577&13034049 \\
$9$&513&1025&20706&326273&5022543&92168924 \\
$10$&1025&2049&61097&1265793&25659185&657904329 \\
$11$&2049&4097&181245&4933121&131347909&4735839173 \\
\end{tabular}
\caption{The number $c$ of homogeneous coordinates for the model ${\bf C}^3/{\mbox{\boldmath $Z$}}_n \times {\mbox{\boldmath $Z$}}_m$.}
\label{table:multiplicity}
\end{center}
\end{table}

The analyses is also applicable to the orbifold ${\bf C}^3/{\mbox{\boldmath $Z$}}_n$.
The McKay quiver of the orbifold ${\bf C}^3/{\mbox{\boldmath $Z$}}_n$ is obtained from that of the orbifold
${\bf C}^3/{\mbox{\boldmath $Z$}}_n \times {\mbox{\boldmath $Z$}}_n$ by the identification \cite{HU},
\begin{equation}
R_{i,j} \equiv R_{i-a, j+1}.
\end{equation}
Thus if we consider configurations of spins which satisfies this identification,
we are able to compute the toric generators and the degeneracy.

\vskip 1cm
\centerline{\large\bf Acknowledgements}

I would like to thank S. Hosono for valuable discussions.

\end{document}